\documentclass[conference]{IEEEtran}
\IEEEoverridecommandlockouts
\usepackage{cite}
\usepackage{amsmath,amssymb,amsfonts}
\usepackage{algorithmic}
\usepackage{graphicx}
\usepackage{textcomp}
\usepackage{xcolor}
\def\BibTeX{{\rm B\kern-.05em{\sc i\kern-.025em b}\kern-.08em
    T\kern-.1667em\lower.7ex\hbox{E}\kern-.125emX}}
\begin{document}

\title{Flat beam plasma wakefield accelerator\\

\thanks{This work was performed with support of the US Dept. of Energy, Division of High Energy Physics, under contract no. DE-SC0017648 and DESC0009914.}
}

\author{\IEEEauthorblockN{Pratik Manwani}
\IEEEauthorblockA{\textit{Department of Physics and Astronomy} \\
\textit{University of California, Los Angeles}\\
Los Angeles, USA \\
pkmanwani@ucla.edu}
\and
\IEEEauthorblockN{Nathan Majernik}
\IEEEauthorblockA{\textit{Department of Physics and Astronomy} \\
\textit{University of California, Los Angeles}\\
Los Angeles, USA \\
nmajernik@g.ucla.edu }
\and
\IEEEauthorblockN{Joshua Mann}
\IEEEauthorblockA{\textit{Department of Physics and Astronomy} \\
\textit{University of California, Los Angeles}\\
Los Angeles, USA \\
jomann@physics.ucla.edu}
\and
\IEEEauthorblockN{Havyn Ancelin}
\IEEEauthorblockA{\textit{Department of Physics and Astronomy} \\
\textit{University of California, Los Angeles}\\
Los Angeles, USA \\
havynancelin@g.ucla.edu}
\and
\IEEEauthorblockN{Yunbo Kang}
\IEEEauthorblockA{\textit{Department of Physics and Astronomy} \\
\textit{University of California, Los Angeles}\\
Los Angeles, USA \\
conrad278@g.ucla.edu}
\and
\IEEEauthorblockN{Derek Chow}
\IEEEauthorblockA{\textit{Department of Physics and Astronomy} \\
\textit{University of California, Los Angeles}\\
Los Angeles, USA \\
dc.7775@gmail.com}
\and
\IEEEauthorblockN{Gerard Andonian}
\IEEEauthorblockA{\textit{Department of Physics and Astronomy} \\
\textit{University of California, Los Angeles}\\
Los Angeles, USA \\
gerard@physics.ucla.edu }
\and
\IEEEauthorblockN{James Rosenzweig}
\IEEEauthorblockA{\textit{Department of Physics and Astronomy} \\
\textit{University of California, Los Angeles}\\
Los Angeles, USA \\
rosen@physics.ucla.edu }
}

\maketitle

\begin{abstract}
Particle beams with highly asymmetric emittance ratios are expected at the interaction point of high energy colliders. 
These asymmetric beams can be used to drive high gradient wakefields in dielectrics and plasma. 
In the case of plasma, the high aspect ratio of the drive beam creates a transversely elliptical blowout cavity and the asymmetry in the ion column creates asymmetric focusing in the two transverse planes. 
The ellipticity of the blowout depends on the ellipticity and normalized charge density of the beam. 
In this paper, simulations are performed to investigate the ellipticity of the wakefield based on the initial driver beam parameters.
The matching conditions for this elliptical cavity are discussed. Example cases for employment using the attainable parameter space at the AWA and FACET facilities are also presented. 
\end{abstract}


\section{Introduction}

Plasma wakefield acceleration (PWFA) using elliptical beams with highly asymmetric emittances, or 'flat beams' requires investigation for future collider applications. 
Flat beams are employed at accelerator facilities like AWA \cite{tianzhe} and are also foreseen at FACET-II \cite{alex}. Flat drive beams yield a blowout cavity that is elliptical in cross section, which leads to asymmetric transverse focusing, and have been proposed for numerous  applications. 
For example, asymmetric beams that drive wakefields in hollow channel plasmas are considered for positron acceleration \cite{zhou}. For colliders, beams with highly asymmetric emittance are expected to mitigate beam-beam effects (beamstrahlung) at the interaction point \cite{chen}. These applications require accurate accounting of the matching conditions and beam loading effects of these asymmetric beams. Both the axisymmetric and approximated elliptical wakefield have linear focusing fields which allow us to match these beams.
It is also important to understand how flat beams behave in plasma afterburner scenarios \cite{manwani1,manwani2}, and compatibility with corresponding betatron diagnostics \cite{monika}. Practical considerations for the development of a flat beam PWFA experiment at the AWA facility, and its potential development at FACET-II, will be presented here.

\section{Linear regime}

\begin{figure}[htbp]
\centerline{\includegraphics[width=\columnwidth]{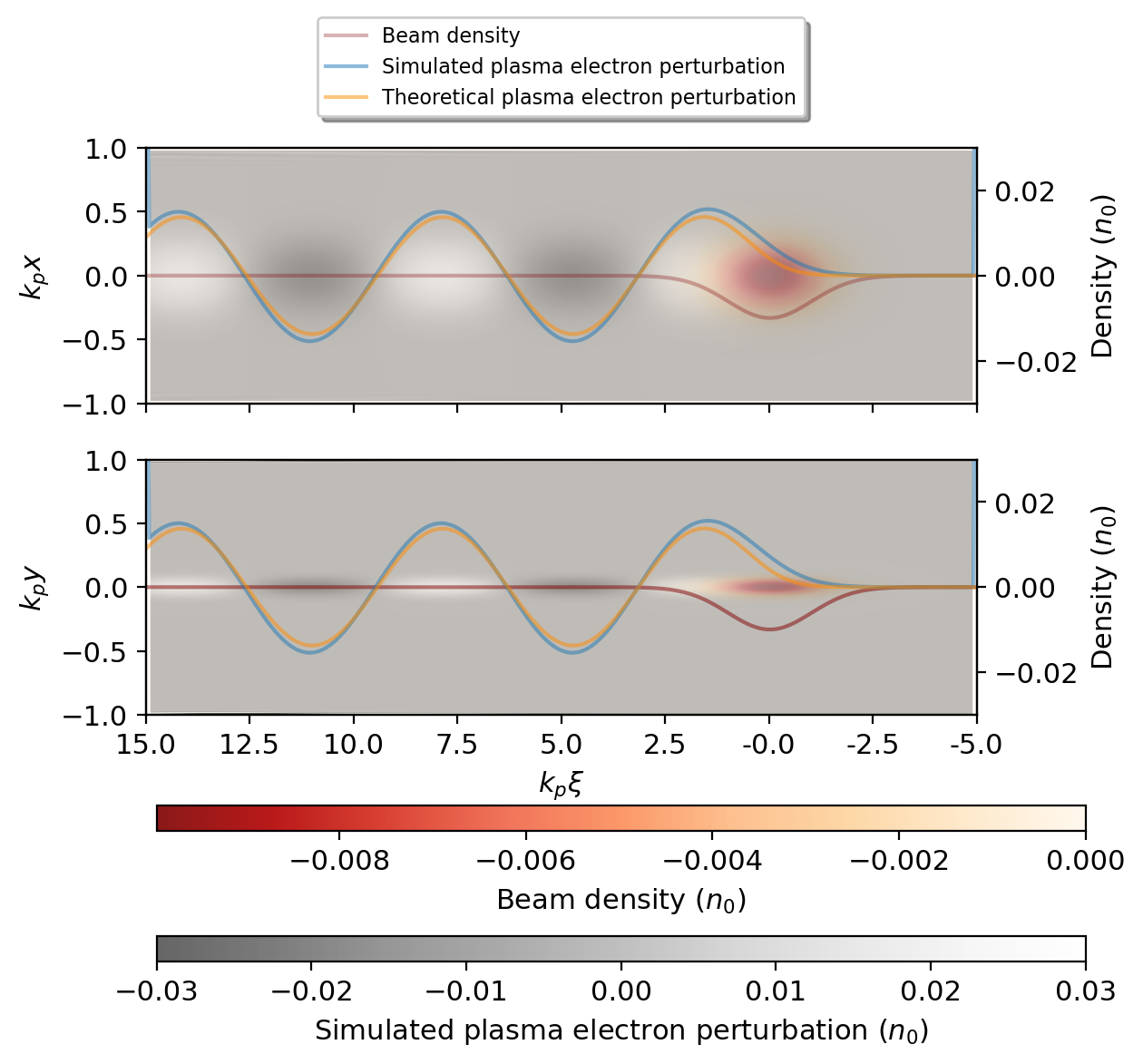}}
\caption{PIC simulations showing the asymmetric wakefield created by an asymmetric driver when the beam density is low ($n_b/n_0 \ll 1$)}
\label{linear}
\end{figure}

We first consider a beam with general form, $n_b = n_{b,0} X(x) Y(y) Z(ct - z)$. If the beam density is small compared to the background plasma density, $n_{b,0}/n_0 \ll 1$ we can use the linearized wake equation, with $\xi = ct - z$, to get the perturbed plasma electron density ($n_1$):

\begin{align}
    \left(\frac{\partial^2}{\partial{\xi}^2} + k_p^2\right) n_1(x,y,\xi) = -k_p^2 n_{b,0} X(x) Y(y) Z(\xi)
\end{align}

where $k_p=\omega_p/c$ is the plasma wavenumber. For a Gaussian beam, this gives

\begin{equation}
\begin{split}
    n_1(x,y,\xi) = -k_p n_{b,0} \exp{\frac{-x^2}{2\sigma_x^2}}\exp{\frac{-y^2}{2\sigma_y^2}} \\\int_\epsilon^\infty \exp{\frac{-\xi'^2}{2\sigma_z^2}} \sin{\left(k_p (\xi - \xi')\right)} d \xi'
    \label{eq}
\end{split}
\end{equation}

where $\sigma_{x,y,z}$ are the spot sizes of the beam. We can simulate the beam plasma interaction in this linear-fluid limit, by using particle in cell (PIC) simulations with the three dimensional fully relativistic PIC code, OSIRIS \cite{osiris}. The simulation results of the interaction in this regime with the corresponding analytical results are shown in Figure \ref{linear}.
The full response of the fields can be found from the density response given in equation \ref{eq} and the electromagnetic wave equations which are given as:

\begin{align}
    (\nabla_\perp^2 - \frac{\omega_p^2}{c^2}) \vec{E} = - e \nabla( \frac{n_1 + n_b}{\epsilon_0}) - \frac{e}{\epsilon_0 c} \frac{\partial n_b}{\partial t}\hat{z} \\
    (\nabla_\perp^2 - \frac{\omega_p^2}{c^2}) \vec{B} = 4 \pi e(\nabla \times n_b \hat{z})
\end{align}

\section{Elliptical blowout}

\begin{figure}[htbp]
\centerline{\includegraphics[width=3.3in
]{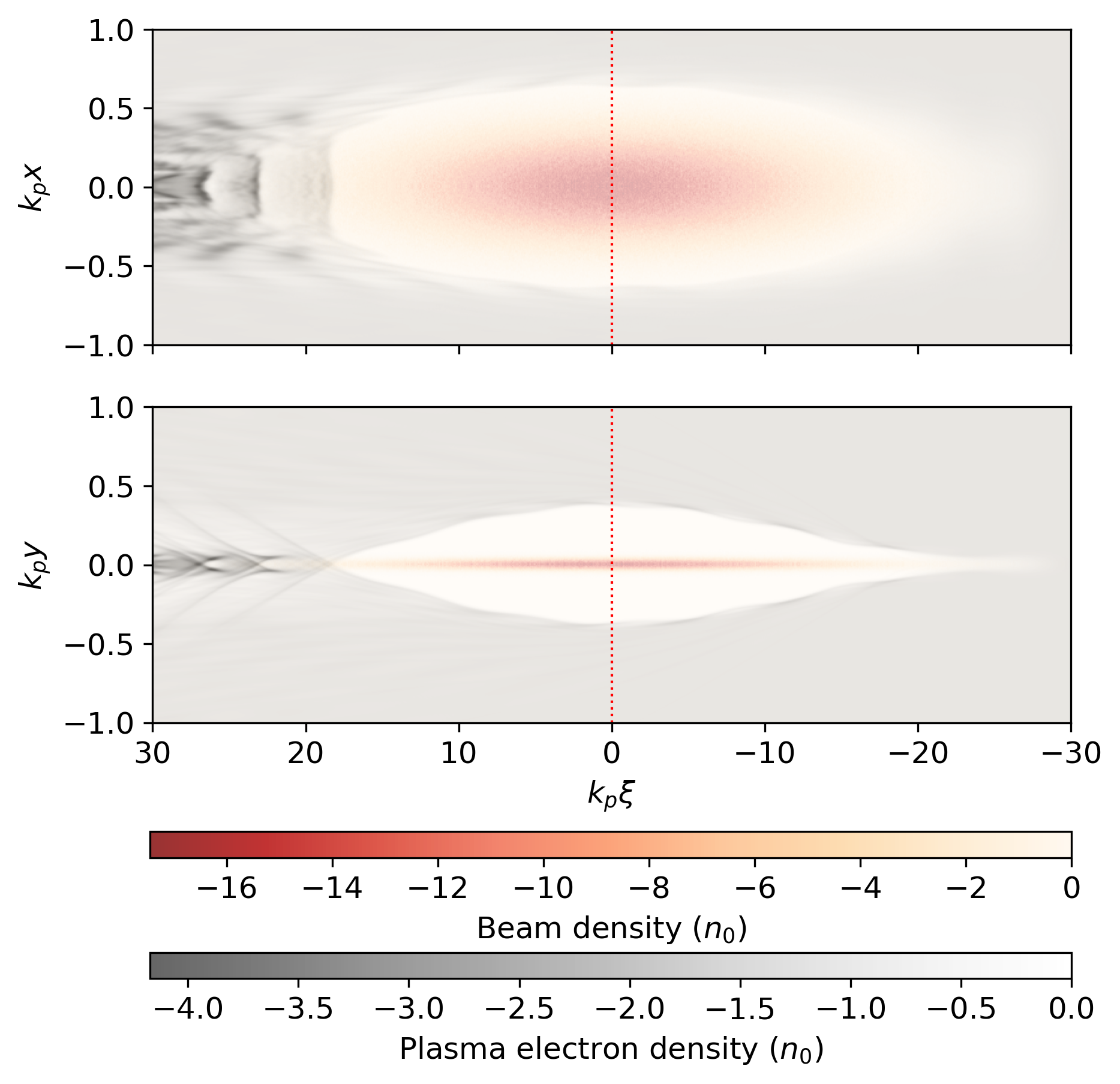}}
\caption{The longitudinal profile of the elliptical blowout cavity formed by an asymmetric (flat) beam ($\sigma_x/\sigma_y = 10$)}
\label{asymmetry}
\end{figure}

\begin{figure}[htbp]
\centerline{\includegraphics[width=3in]{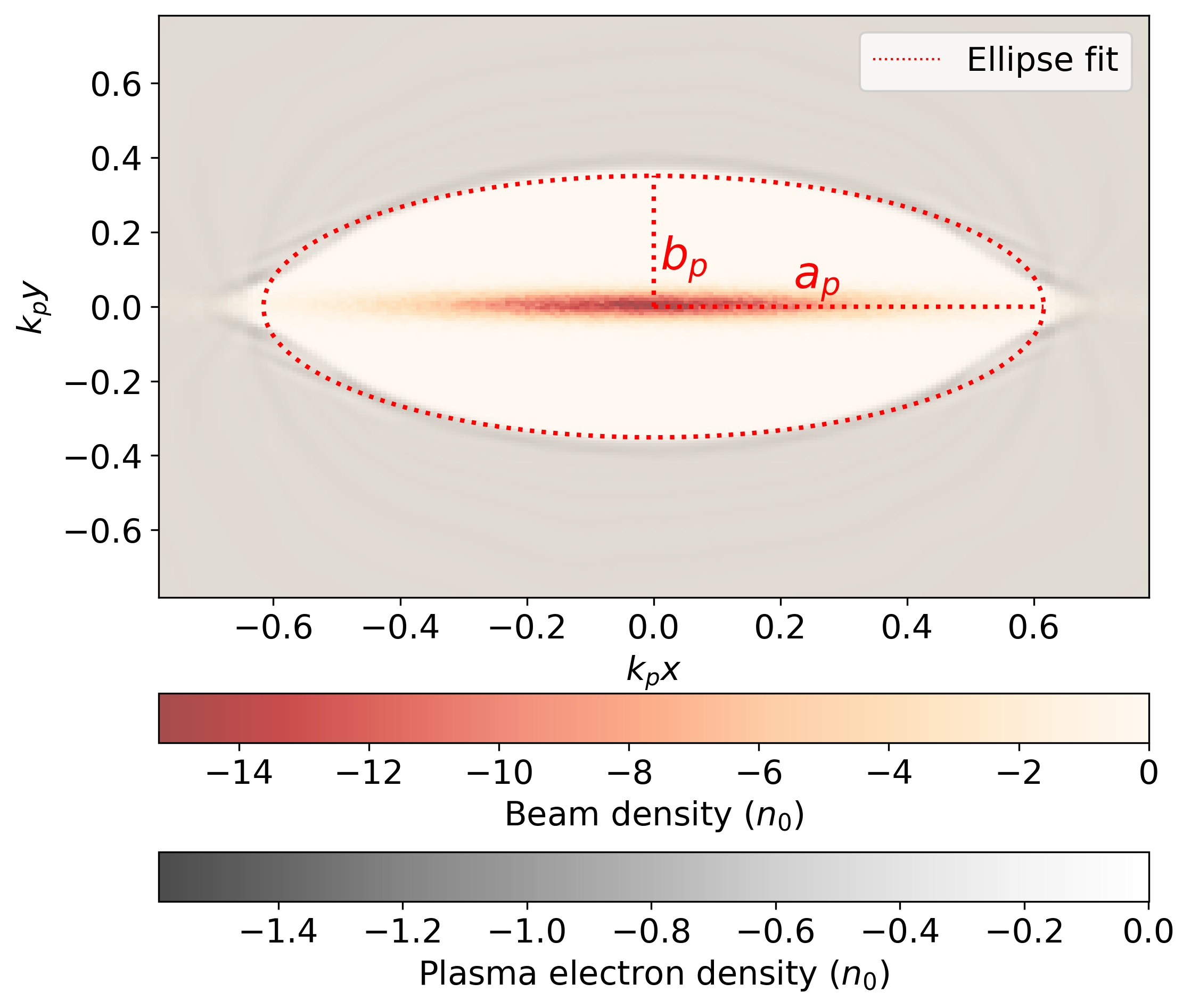}}
\caption{The transverse profile of the center of the elliptical blowout cavity formed by an asymmetric (flat) beam ($\sigma_x/\sigma_y = 10$)  The radii of the ellipse are shown as $a_p$ and $b_p$}
\label{ellipse}
\end{figure}

For high beam densities, the plasma electron trajectories cross and there is a formation of an elliptical blowout sheath. We simulate this by using particle in cell (PIC) simulations with the three dimensional quasi-static PIC code, QuickPIC \cite{quickpic}. The simulation parameters are listed in table \ref{sim}. The transverse asymmetry of the wakefield produced by the asymmetric beam can be seen in Figure \ref{asymmetry}. The single density spike that is observed in the axisymmetric case is replaced with a cluster of electrons that can be seen as a  line of high density behind the blowout region due to the ellipsoidal shape of the wake. This elliptical cavity created by the evacuated plasma electrons and its radii ($a_p$ and $b_p$) are shown in Figure \ref{ellipse}. Inside this ion column ($\frac{x^2}{a_p^2} + \frac{y^2}{b_p^2} < 1$), the wake equations can be written as:

 \begin{align}
      \phi (\xi,\alpha_b,\alpha_p) = \phi_0 (\xi) + \phi_p (\xi,\alpha_p) + \phi_b (\xi,\alpha_b) \\
      A_z (\xi, \alpha_b) = A_{z0} (\xi) + c \Phi_b (\xi, \alpha_b) \\
      F_x = \partial_x \phi_p + (1-v_z)\partial_x \phi_b + (1-v_z) \partial_\xi A_x - v_y B_z \\
      F_y = \partial_y \phi_p + (1-v_z)\partial_y \phi_b + (1-v_z) \partial_\xi A_y + v_x B_z
  \end{align}

Here, the parameter $\alpha_b$ is defined as the beam ellipticity of a beam with spatial extent of $a$ and $b$ in the $x$ and $y$ direction respectively, and $\alpha_p$ is defined as the ellipticity of the transverse cross section of the elliptical cavity. In addition, $\phi$ denotes the scalar potential, $\vec{A}$ denotes the vector potential, and $F_x$ and $F_y$ denote the transverse forces inside the blowout cavity.

\begin{table}[htbp]
\caption{Simulation parameters}
\begin{center}
\begin{tabular}{|c|c|c|}
\hline
%\textbf{Table}&\multicolumn{3}{|c|}%{\textbf{Table Column Head}} \\
%\cline{2-4} 
%\textbf{Head} 
\textbf{\textit{Parameter}}& \textbf{\textit{Value}}& \textbf{\textit{Unit}} \\
\hline
        Beam density, $n_b$ & 15 & $n_0$ \\
        Energy, $E_b$ & 50 & MeV \\
        $\sigma_z$ & 10 & $k_p^{-1}$ \\
        $\sigma_x$, $\sigma_y$ & 0.25,0.025 & $k_p^{-1}$\\
        $\epsilon_x$, $\epsilon_y$ & 200, 2 & $\mu$m-rad\\
        %Peak current, $I_{pk}$ & 960 A\\
        Plasma density, $n_0$ & 7 $\times$ $10^{13}$ & cm$^{-3}$ \\
\hline
%\multicolumn{4}{l}{$^{\mathrm{a}}$Sample of a Table footnote.}
\end{tabular}
\label{sim}
\end{center}
\end{table}

As a starting point, we can get the ellipticity at the center of the blowout cavity using the long beam ($r \ll \gamma \sigma_z$) approximation where we neglect the longitudinal variation of the fields and also neglect the effects of the plasma electron velocity. Under this approximation:

\begin{align}
    \partial_{x,y} \phi_p  = -\partial_{x,y} \phi_b
\end{align}

The potential inside an ellipsoid is quadratic with respect to the coordinate, which leads to linear fields. The transverse fields in the elliptical blowout structure can be calculated analytically by approximating it as an infinite long cylinder of ions, similar to the axisymmetric case but with an elliptical cross section and are given by:

\begin{align}
    E_{x,p} = \frac{e n_0 b_p x}{\epsilon_0 (a_p+b_p)} = \frac{e n_0 x}{\epsilon_0 (1+\alpha_p)}\\
    E_{y,p} = \frac{e n_0 a_p y}{\epsilon_0 (a_p+b_p)} = \frac{e n_0 \alpha_p y}{\epsilon_0 (1+\alpha_p)}
\end{align}

where $a_p$ and $b_p$ are the elliptical cross section's semimajor and semiminor axes respectively, and the ellipticity $\alpha_p$ is defined as the ratio $a_p/b_p$.

Outside the beam (assuming a uniform density beam with no longitudinal variation), the fields are given as \cite{parzen}:

\begin{align}
    E_{x,b} &= \frac{e n_b a b}{\epsilon_0 (x + (x^2 + b^2 - a^2)^{1/2})} \\
    E_{y,b} &= \frac{e n_b a b}{\epsilon_0 (y + (y^2 + a^2 - b^2)^{1/2})}
\end{align}

Equating these fields on the boundaries, we get:

\begin{align}
    a_p = a \sqrt{\frac{\alpha_b^2-1}{2 \alpha_b^2}} \sqrt{1 + \sqrt{1 + \frac{4 \alpha_b^2 n_b^2 }{(\alpha_b^2-1)^2}}} \\
    b_p = a \sqrt{\frac{\alpha_b^2-1}{2 \alpha_b^2}} \sqrt{-1 + \sqrt{1 + \frac{4 \alpha_b^2 n_b^2 }{(\alpha_b^2-1)^2}}}
\end{align}

The ellipticity is then given by:

\begin{align}
    \alpha_p = \frac{a_p}{b_p} = \left(\frac{\sqrt{\frac{4 \alpha_b^2 n_b^2 }{\left(\alpha_b^2-1\right)^2}+1}+1}{\sqrt{\frac{4 \alpha_b^2 n_b^2}{\left(\alpha_b^2-1\right)^2}+1}-1}\right)^\frac{1}{2}
\end{align}

\begin{figure}[htbp]
\centerline{\includegraphics[width=3.3in]{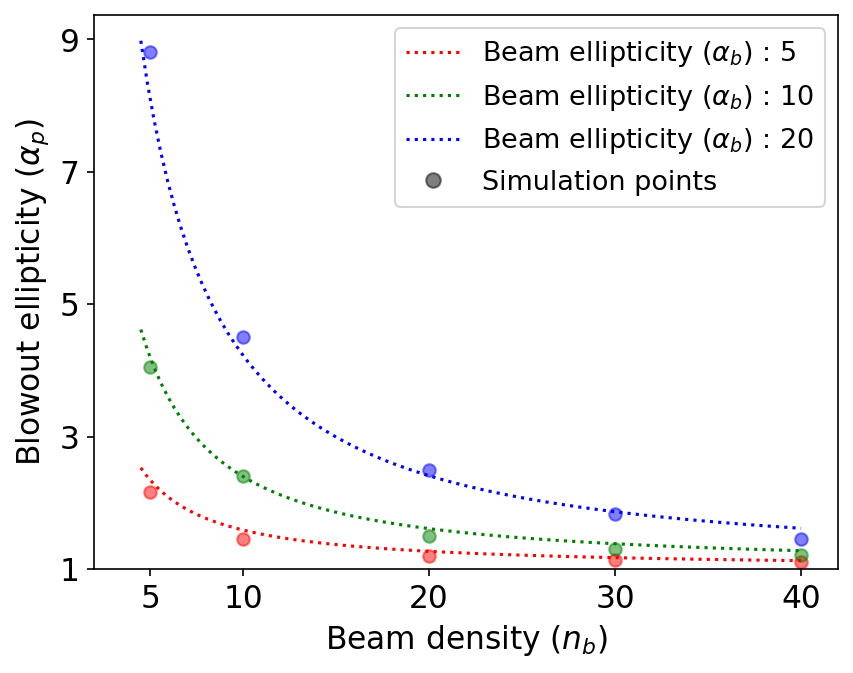}}
\caption{Comparison of analytical and simulated blowout ellipticities at $\xi=0$ created by long Gaussian beams ($k_p \sigma_z$=10). The dotted curves represents the analytical solution for the blowout ellipticity for a given beam ellipticity ($\alpha_b$), while the markers correspond to the simulation points obtained at initialization of the plasma wakefields}
\label{match}
\end{figure}

The non-linearity of the wake along with the ellipticity reduces with an increase in beam density. Simulations using long Gaussian beams agree with this equation at initialization (To make the linear charge density constant, we scale the beam density by a factor of 2). Specifically, the asymmetric ellipse fits were obtained from a least squares fit performed on the curve defined by the border between the blowout and the surrounding neutral plasma.
This border is defined where the normalized plasma density
falls below 1/e. This method of fitting works well when the blowout is stronger as the sheath is more distinct at higher beam densities. There is good agreement between the predicted ellipticity and the simulated ellipticity at the center of the blowout cavity and the results are shown in Figure \ref{match}. The approximation of neglecting the plasma electron velocity does not hold for high beam densities, and will need to be taken into account to get more accurate results. 

\section{Matching conditions}

The focusing forces can be directly derived from the ellipticity of the blowout cavity by modeling it as an infinite elliptical cylinder as is done for the axisymmetric case. These linear forces can be used to match the beam into the blowout cavity by countering the divergence of the beam. A mismatch will lead to characteristic beating of the beam envelope, causing an increase in the transverse emittance due to phase space dilution \cite{betatron,manwani2} .The matched beta functions for the beam envelope are:
%\begin{align}
%    K_x^2 = \frac{1}{\gamma m _e c^2}\frac{F_{x}}{x} = \frac{n_0 e^2}{\gamma m _e c^2\epsilon_0(1 + \alpha_p)} \\
%    K_y^2 = \frac{1}{\gamma m _e c^2}\frac{F_{y}}{y} = \frac{n_0 e^2  \alpha_p}{\gamma m _e c^2 \epsilon_0(1 + \alpha_p)}
%\end{align}

\begin{align}
    k_{\beta,x} = \sqrt{\frac{1}{\gamma m _e c^2}\frac{eE_{x}}{x}} = \sqrt{\frac{n_0 e^2}{\gamma \epsilon_0m_ec^2(1+\alpha_p)}} \\
    k_{\beta,y} = \sqrt{\frac{1}{\gamma m _e c^2}\frac{eE_{y}}{y}} = \sqrt{\frac{n_0 \alpha_p e^2}{ \gamma \epsilon_0 m_ec^2(1+\alpha_p)}}
\end{align}

The spot size of the beam is then given by $\sigma_{x,m}$ = $\sqrt{\beta_x \epsilon_x}$ and $\sigma_{y,m}$ = $\sqrt{\beta_y \epsilon_y}$. The ellipticity of the matched beam, $\alpha_{b, m}$, is then given by:

\begin{align}
    \alpha_{b,m} = \frac{\sigma_{x,m}}{\sigma_{y,m}} = \sqrt{\frac{\epsilon_x}{\epsilon_y}\frac{\beta_x}{\beta_y}} = \sqrt{\frac{\epsilon_x}{\epsilon_y}\frac{k_{\beta,y}}{k_{\beta,x}}} = \sqrt{\frac{\epsilon_x}{\epsilon_y}\sqrt{\alpha_{p,m}}} 
    \label{matching_eq}
\end{align}

The emittance ratio is

\begin{align}
    \frac{\epsilon_x}{\epsilon_y} = \alpha_b^2 \left(\frac{\sqrt{\frac{4 \alpha_b^2 n_b^2 }{\left(\alpha_b^2-1\right)^2}+1}-1}{\sqrt{\frac{4 \alpha_b^2 n_b^2}{\left(\alpha_b^2-1\right)^2}+1}+1}\right)^\frac{1}{4}
\end{align}

%\begin{align}
%    \alpha_{b,m} = \frac{\sigma_{x,m}}{\sigma_{y,m}} = \sqrt{\frac{\epsilon_x}{\epsilon_y}\sqrt{\alpha_{p,m}}} 
%    \label{matching_eq}
%\end{align}

This result shows that the emittance needed to match the beam envelope to the focusing forces needs to account for the ellipticity of the plasma column. The blowout ellipticity and concomitantly, the emittance ratio required to match the beam will vary along the length of the blowout cavity and this will be discussed in a subsequent paper.

\iffalse
\begin{table}[htbp]
\caption{Simulation parameters}
\begin{center}
\begin{tabular}{|c|c|c|c|}
\hline
\textbf{Table}&\multicolumn{3}{|c|}{\textbf{Table Column Head}} \\
\cline{2-4} 
\textbf{Head} & \textbf{\textit{Table column subhead}}& \textbf{\textit{Subhead}}& \textbf{\textit{Subhead}} \\
\hline
copy& More table copy$^{\mathrm{a}}$& &  \\
\hline
\end{tabular}
\label{tab1}
\end{center}
\end{table}
\fi

\section{Experimental considerations}

The AWA facility is an ideal location to experimentally investigate key features of the flat-beam driven PWFA. The experiment at AWA \cite{manwani1} involves using the asymmetric beam to drive a weak blowout in a plasma created by a capillary discharge. The capillary discharge plasma source, commissioned at UCLA, has a multi-decade range of densities, which will allow exploration of the varying regimes described in this paper. The condition of weak blowout creates head erosion in the drive beam which decreases the beam density and leads to the increase in blowout ellipticity. This dynamicity is also reflected in the beam matching condition  and an optimal solution to this problem require consideration of many factors.  
In the other scenario of strong blowout, beams that will be available at FACET-II would have significantly higher beam densities at the interaction point. At these high densities, an asymmetric beam will create an axisymmetric blowout. Preliminary simulations
with the asymmetric FACET-II beam show the bifurcation
of the beam after long propagation distances, in which the
beam splits into two equal halves. This could be a potential violation of the quasi-static condition implied in QuickPIC and this  extreme beam density case would need to be benchmarked with a full 3D PIC code but the asymmetric nature of the bunches makes it computationally intensive.

\section{Conclusion}

The structure of the plasma columns formed by elliptical beams determines the focusing forces of the wakefield which governs the dynamics of the beam. This makes it integral to understand the formation and ellipticity of the plasma column for plasma wakefield experiments and plasma afterburner scenarios which employ asymmetric beams.  The results shown here provide information on the ellipticity and emittance requirements for matching the beam inside the elliptical blowout cavity. 
%This paper lays the groundwork for this but further work will be required to sufficiently characterize the asymmetric wakefield. 
The limiting case for the flat beam has been investigated theoretically \cite{stas} and work is ongoing to 
understand and generalize the properties of this asymmetric beam plasma interaction. The inequality in the focusing forces in the two transverse planes offers novel opportunities, one of which is the creation of an asymmetric plasma lens which is a subject of further study.
%#\section{Conclusion}
\section*{Acknowledgment}

This work was performed with the support of the US Department of Energy under Contract No. DE-SC0017648 and DESC0009914. This work used resources of the National Energy Research Scientific Computing Center (NERSC), a U.S. Department of Energy Office of Science User Facility located at Lawrence Berkeley National Laboratory, operated under Contract No. DE-AC02-05CH11231.

%\section*{References}
.

\end{document}